# Economic Models for Management of Resources in Grid Computing

Rajkumar Buyya, Heinz Stockinger‡, Jonathan Giddy, and David Abramson

| CRC for Enterprise Distributed Systems Technology | ‡CERN, European Organization for Nuclear Research |
| --- | --- |
| School of Computer Science and Software Engineering | CMS Experiment, Computing Group, Database Section |
| Monash University, Melbourne, Australia | CH-1211 Geneva 23, Switzerland |

Email: {rajkumar, jon, davida}@csse.monash.edu.au, Heinz.Stockinger@cern.ch

**Abstract**

The accelerated development in Grid and peer-to-peer computing has positioned them as promising next generation computing platforms. They enable the creation of *Virtual Enterprises* (VE) for sharing resources distributed across the world. However, resource management, application development and usage models in these environments is a complex undertaking. This is due to the geographic distribution of resources that are owned by different organizations. The resource owners of each of these resources have different usage or access policies and cost models, and varying loads and availability. In order to address complex resource management issues, we have proposed a computational economy framework for resource allocation and for regulating supply and demand in Grid computing environments. The framework provides mechanisms for optimizing resource provider and consumer objective functions through trading and brokering services. In a real world market, there exist various economic models for setting the price for goods based on *supply-and-demand* and their value to the user. They include commodity market, posted price, tenders and auctions. In this paper, we discuss the use of these models for interaction between Grid components in deciding resource value and the necessary infrastructure to realize them. In addition to normal services offered by Grid computing systems, we need an infrastructure to support interaction protocols, allocation mechanisms, currency, secure banking, and enforcement services. Furthermore, we demonstrate the usage of some of these economic models in resource brokering through Nimrod/G deadline and cost-based scheduling for two different optimization strategies on the World Wide Grid (WWG) testbed that contains resources located on five continents: Asia, Australia, Europe, North America, and South America.

## 1. Introduction

Peer-to-Peer (P2P) Grid computing has emerged as a new paradigm for solving large-scale problems in science, engineering, and commerce [1][5]. The P2P and Grid technologies enable the creation of *Virtual Enterprises* (VE) for resource sharing by logically coupling millions of resources (e.g., SETI@Home [23]) scattered across multiple geographically distributed organizations, administrative domains, and policies. They comprise heterogeneous resources (PCs, Work statations, clusters, and supercomputers), fabric management systems (single system image OS, queuing systems, etc.) and policies, and applications (scientific, engineering, and commercial) with varied requirements (CPU, I/O, memory, and/or network intensive). The users, *producers* also called resource owners and *consumers* have different goals, objectives, strategies, and demand patterns. More importantly both resources and end-users are geographically distributed with different time zones. In managing such complex Grid environments, traditional approaches to resource management that attempt to optimize system-wide measure of performance cannot be employed. Traditional approaches use centralized policies that need complete state information and a common fabric management policy, or decentralized consensus based policy. Due to the complexity in constructing successful Grid environments, it is impossible to define an acceptable system-wide performance matrix and common fabric management policy [13].

In [2][3][4][5], we proposed and explored the usage of an economics based paradigm for managing resource allocation in Grid computing environments. The economic approach provided a fair basis in successfully managing decentralization and heterogeneity that is present in human economies. Competitive economic models provide algorithms/policies and tools for resource sharing/allocation in Grid systems. The models can be based on bartering/exchange or prices. In the *bartering-based model*, all participants need to own resources and trade/share resources by exchanges (e.g., storage space for CPU time). In the *price-based model*, the resources are priced, based on the demand, supply, value, and the wealth of economics systems.

The resource management systems need to provide mechanisms and tools that facilitate the realization of goals for both service providers (resource owners) and consumers. The resource consumers need a *utility model*—how do consumers demand resources and what are their preference parameters—and *brokers* that automatically generate strategies for choosing providers as per user requirements and manage all issues associated with application execution. The service providers need tools/mechanisms for *price generation schemers* so as to increase system utilization and *protocols* that help them to offer competitive services. For the market to be competitive and healthy, coordination mechanisms are required that help in reaching equilibrium price—the market price at which the supply of a service equals the quantity demanded.

Most of the related work in Grid computing dedicated to scheduling problems and resource management uses a "conventional style" where a scheduling component decides which jobs can be executed at which site based on certain cost functions (Globus [6], Legion [7], Condor [28], AppLeS [26], Netsolve [27], Punch [25]). Such cost functions are often very system-centric and are not driven by user Quality of Service (QoS) parameters such as access price and service delivery timeframe. Most systems treat resources as if they all cost the same price even when this is not the case in reality. The end user does not want to pay the highest price but wants to negotiate a particular price based on the demand, value, priority, and available budget. In an economics approach, the scheduling decision is not done statically by a single scheduling entity but directed by the end users requirements. Whereas a conventional cost model often deals with software and hardware costs for running applications, the economic model primarily charges the end user for services that they consume based on the value they derive from it. Trading based on the demand of users and the available resources is the main driver in the competitive, economic market model. Thus, we stress that a single user is in competition with other users and resource owners with Grid service providers.

The main contribution of this paper is to provide economic models, system architecture, and policies for resource management in Grid environments. Currently, the Grid user community and the Grid technology are still rather new and not well accepted and established in commercial Grid settings. However, we believe the Grid can become established in such settings by providing incentive for both consumers and resource providers for being part of the Grid. Since the Grid uses the Internet as a carrier for providing remote services, it is well positioned to create a computational ecology, cooperative problem solving environment, and means for sharing digital contents and resources in a seamless manner. Up to now, the idea of using Grids for solving large computationally intensive applications has been more or less restricted to the scientific community. However, even if, in the scientific community, the pricing aspect seems to be of minor importance, there are still funding agencies that have to support the hardware and software infrastructure for Grids. Economic models help them in management and evaluation of resource allocations to user communities.

## 2. Players in the Grid Marketplace

The two key players driving the Grid marketplace, like in the conventional marketplace, are: *Grid Service Providers* (GSPs) providing the traditional role of *producers* and *Grid Resource Brokers* (GRBs) representing *consumers*. The Grid computing environments provide necessary infrastructure including security, information, transparent access to remote resources, and information services that enable us to bring these two entities together. Consumers interact with their own brokers for managing and scheduling their computations on the Grid. The GSPs make their resources Grid enabled by running software systems (such as Globus or Legion) along with *Grid resource Trading Services/Servers* (GTS) to enable resource trading and execution of consumer requests directed through GRBs. The interaction between GRBs and GSPs during resource trading is mediated through a *Grid Market Directory* (GMD) (see Figures 1 to 5). They use various economic models or interaction protocols driven by a Grid Marketplace for deciding service access price. These protocols are discussed in Section 3. The Grid Architecture for Computational Economy (GRACE) discussed in [5] provides detailed discussion on the role played by GSPs, GRBs, and GMDs and compares them to actual Grid components and implementations.

As in the conventional marketplace, the users' community (GRBs) represents the demand, whereas the resource owners' community (GSPs) represents the supply in our economic Grid model. In the economic model, we put emphasis on the user community and how they can influence the pricing of Grid resources via their brokers. Prices are only one factor in the model but play an important role when resources are only used and not bought by the users, as for real world services. The service providers (GSPs) and consumers (GRBs) interact in a competitive market environment for resource trading and service access.

In the literature, the term *economics* is defined as the branch of social science that deals with the production and distribution and consumption of goods and services and their management [17]. Numerous economic theories including microeconomic and macroeconomic principles have been proposed in the literature. Some of the commonly used economic models for selling goods and services can be employed as service price negotiation protocols in Grid computing, include:

- Commodity Market Model (case systems: Mungi [18], Enhanced MOSIX [19], and Nimrod/G [1][2])
- Posted Price Models (case system: Nimrod/G)
- Bargaining Model (case systems: Mariposa [8] and Nimrod/G,)
- Tendering/Contract-Net Model (case system: Mariposa [8])
- Auction Model (case systems: Spawn [20] and Popcorn [21])
- Bid-based Proportional Resource Sharing Model (case system: Rexec/Anemone [22])
- Community/Coalition/Bartering Model (case systems: Condor, SETI@Home [23], Mojo Nation [24])
- Monopoly, Oligopoly (Nimrod/G broker—it can still choose between resources offered at different price)

Derivatives like options and futures might also be interesting to consider. They have a high risk but allow users to hedge against price increases. The use of the above models in Grid computing is discussed in the next section. Typically, in a Grid marketplace, the resource owners, i.e. Grid Service Providers, and users can use any one or more of these models or even combinations of them in meeting their objectives [5]. Both have their own expectations and strategies for being part of the Grid. In this Grid economy, resource consumers adopt the strategy of solving their problems at low cost within a required timeframe and resource providers adopt the strategy of obtaining best possible return on their investment. The resource owners try to maximize their resource utilization by offering a competitive service access cost in order to attract consumers. The resource consumers have an option of choosing the providers that best meet their requirements.

Any of them (GRBs or GSPs) can initiate a resource trading in a market like environment and participate in the negotiation depending on their objectives. GRBs can invite bids from a number of GSPs and select those that offer lowest services costs and meet their requirements that are driven by user deadline and budget. GSPs can also invite bids in an auction like environment and offer services to highest bidder as long as its objectives are met. Both of them, GSPs and GRBs, have their own utility functions that have to be satisfied and maximized. The consumers perform a cost-benefit analysis depending on the deadline (by which results are required) and budget available (the amount of money the user is willing to invest for solving the problem). The resource owners decide their pricing based on various factors [5]. They may charge different prices for different users for the same service or it can vary depending on the specific user demands. Resources may have different prices based on the environmental influences.

In a Grid marketplace, Grid brokers (note that in a Grid environment each user has his own broker as his agent) may have different goals (e.g., different deadlines and budgets), and each broker tries to maximize its own good without concern for the global good. Such self-interest naturally encourages negotiations among independent businesses or individuals. This needs to be taken into consideration in building automated negotiation infrastructure. In a *cooperative distributed computing or problem-solving environment* (like cluster computers), the system designers impose an *interaction protocol* (possible actions to take at different points) and a *strategy* (a mapping from one state to another and a way to use the protocol). This model aims for global efficiency as nodes cooperate towards a common goal. On the other hand, in Grid systems, brokers and GSPs are provided with an interaction protocol, but each broker and GSP chooses the best strategy for itself (like in multi-agent systems), which cannot be imposed from outside. Therefore, the negotiation protocols need to be designed using a *non-cooperative, strategic* perspective. In this case, the main concern is what social outcomes follow given a protocol, which *guarantees that each broker/GSP's desired local strategy is best for that broker/GSP and hence the broker/GSP will use it.*

Various criteria used for judging effectiveness of a market model are [15]:

- Social welfare (global good of all)
- Pareto efficiency (global perspective)
- Individual rationality (better off by participating in negotiation)
- Stability (mechanisms that cannot be manipulated, i.e., behave in the desired manner)
- Computational efficiency (protocols should not consume too much computation time)
- Distribution and communication efficiency (communication overhead to capture a desirable global solution).

## 3. Economic Models in a Grid Context

In the previous section we identified a few popular models that are used in human economies. In this section we discuss the use of those economic models and propose a Grid architecture for realizing them. The discussion on realizing negotiation protocols based on different economic models is kept as generic as possible. This ensures that our proposed architecture is free from any specific implementation and provides a general framework for any other Grid middleware and tools developers. Particular emphasis will be placed on heuristics that Grid resource brokers (*G-Brokers*) can employ for establishing service price depending on their customers' requirements. The service providers publish their services through the Grid market directory. They use Grid trading services' declarative language for defining cost specification and their objectives such as access price for various users for different times and durations along with possibilities of offering discounts to attract users during off-peak hours. The Grid trading server can employ different economic models in providing services. The simplest would be a commodity model wherein the resource owners define pricing strategies including those driven by the demand and resource availability. The GTS can act as auctioneer if Auction-based model is used in deciding the service access price or an external auctioneer service can be used (Fig. 6).

For each of the economic models, firstly, the economic model theory, its parameters and influences are discussed and then a possible solution is given for a current Grid environment and how they can be mapped to existing Grid tools and architectures or what needs to be extended. In the classical economic theory there are different models for specific environmental situations and computing applications. Since the end-user interaction is the main interest of this paper, we point out possible interactions with the broker.

### *3.1. Commodity Market (Flat or Supply-and-Demand Driven Pricing) Model*

In the commodity market model, resource owners specify their service price and charge users according the amount of resource they consume. The pricing policy can be derived from various parameters and can be *flat* or *variable* depending on the resource *supply and demand*. In general, services are priced in such a way that supply and demand equilibrium is maintained. In the *flat price model*, once pricing is fixed for a certain period, it remains the same irrespective of service quality. It is not significantly influenced by the demand, whereas in a *supply and demand* model prices change very often based on supply and demand changes. In principle, when the demand increases or supply decreases, prices are increased until there exists equilibrium between supply and demand.

Pricing Schemes in a Commodity Market Model can be based on:
- Flat fee
- Usage Duration (Time)
- Subscription
- Demand and Supply-based [11]

As in the real world, GSPs can publish their prices through yellow pages like Grid market directory (GMD) service (see Figure 1). The resource owners define parameters that GTS can use when providing service prices. A simple price specification may contain the following parameters.

```
{
    . . .
    consumer_id                    // this can be same Grid-ID
    peak_time_price                // 9am-6pm: office hours on working days
    lunch_time_pirce               // (12.30-2pm)
    offpeak_time_price             // (6pm-9am),
    discount_when_lightly_loaded   // if_load is less than 50% at any time
    raise_price_high_demand        //  % raise price if average load is above 50%
    price_holiday_time             //  during holidays and week ends!
    . . .
}
```

Consumers can be charged for access to various resources including CPU cycles, storage, software, and network. The users compose their application using higher-level Grid programming languages. For example, in our Nimrod problem-solving environment we provide a declarative programming language for composing parameter sweep applications and defining application and user requirements such as deadline and budget [1]. The resource broker (working for the user) can carry out the following steps for executing applications:
  a. The broker identifies resource providers
  b. It identifies suitable resources and establishes their prices (GMD and GTS)
  c. It selects resources that meet objectives (lower cost and meet deadline requirements). It uses heuristic techniques while selecting resources and mapping jobs to resources.
  d. It uses them and pays them as agreed.

The actual realization of the above steps is skipped and left to the implementations (e.g., our Nimrod/G resource broker performs all these steps in finer details and supports deadline and budget based scheduling [1] [2][4]).

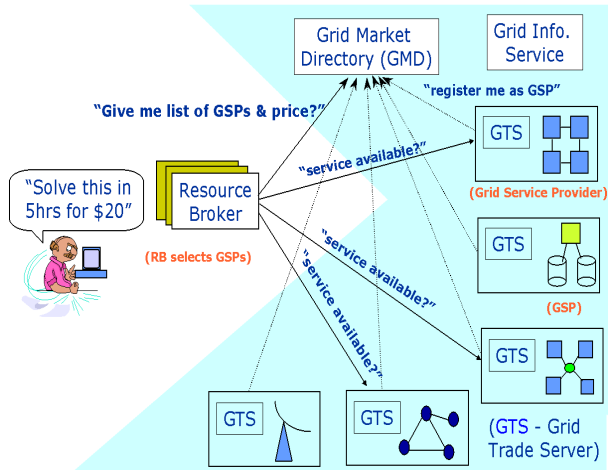 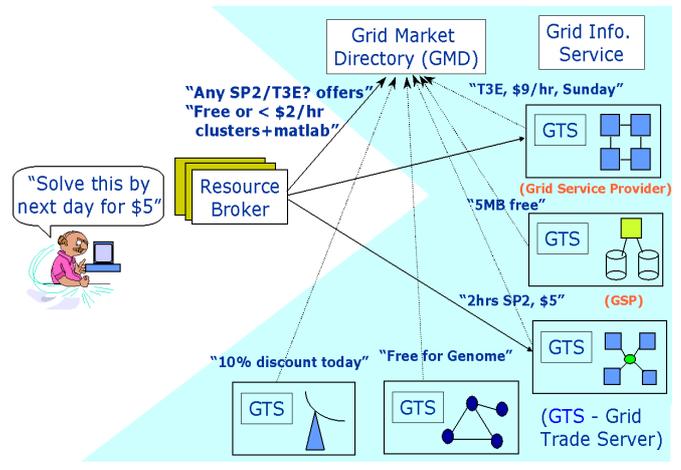

**Figure 1:** Interaction between GSPs and users in a commodity market Grid for resource trading.

**Figure 2:** Posted price model and resource trading in a computational market environment.

### 3.2. Posted Price Model

The posted price model is similar to the commodity market model, except that it advertises special offers (see Figure 2) in order to attract (new) consumers to establish market share or motivate users to consider using cheaper slots. In this case, brokers need not negotiate directly with GSPs for price, but use posted prices as they are generally cheaper compared to regular prices. The posted-price offers will have usage conditions, but they might be attractive for some users. For example, during holiday periods, demand for resources is likely to be limited and GSPs can post tempting offers or prices aiming to attract users to increase resource utilization. The activities that are specifically related to the posted-price model in addition to those related to commodity market model are:

a. Resource/Grid Service Providers (GSPs) posts their service offers and conditions etc. in Grid Market Directory.
b. Broker looks at GMD to identify if any of these posted services are available and fits its requirements
c. Broker enquires (GSP) for availability of posted services.
d. Other steps are similar to those pointed out in commodity market model.

### 3.3. Bargaining Model

In the previous models, the brokers pay access prices, which are fixed by GSPs. In the bargaining model, resource brokers bargain with GSPs for lower access price and higher usage duration. Both brokers and GSPs have their own objective functions and they negotiate with each other as long as their objectives are met. The brokers might start with a very low price and GSPs with a higher price. They both negotiate until they reach a mutually agreeable price (see Figure 3) or one of them is not willing to negotiate any further. This negotiation is guided by user requirements (e.g., deadline is too relaxed) and brokers can take risk and negotiate for cheaper prices as much as possible and can discard expensive machines. This might lead to lower utilization of resources, so GSPs might be willing to reduce the price instead of wasting resource cycles. Brokers and GSPs generally employ this model when market *supply-and-demand* and service prices are not clearly established. The users can negotiate for a lower price with promise of some kind favour or using GSPs services even in the future.

### 3.4. Tender/Contract-Net Model

Tender/Contract-Net model is one of the most widely used models for service negotiation in a distributed problem-solving environment [14]. It is modeled on the contracting mechanism used by businesses to govern the exchange of goods and services. It helps in finding an appropriate service provider to work on a given task. Figure 4 illustrates the interaction between brokers and GSPs in their bid to meeting their objectives. A user/resource broker asking for a task to be solved is called the *manager* and resource that might be able to solve the task is called potential *contractor*. From a manager's perspective, the process is:

1. Consumer (Broker) announces its requirements (using deal template) and invites bids from GSPs.
2. Interested GSPs evaluate the announcement and respond by submitting their bids
3. Broker evaluates and awards the contract to the most appropriate GSP(s)
4. The broker and GSP communicate privately and use the resource (R)

The contents of the deal template used for work announcement include, addressee (user), eligibility requirements specifications (for instance, Linux, x86arch, and 128MB memory), task/service abstraction, optional price that the user is willing to invest, bid specification (what should offer contain), expiration time (deadline for receiving bids).

From a contractor's/GSP perspective, the process is:
1. Receive tender announcements/advertisements (say in GMD)
2. Evaluate service capability
3. Respond with bid
4. Deliver service if bid is accepted
5. Report results and bill the broker/user as per the usage and agreed bid.

The advantage of this model is that if the GSP is unable to provide a satisfactory service or deliver a solution, the Grid resource broker can seek other GSPs for the service. This protocol has certain disadvantages. A task might be awarded to a less capable GSP if a more capable GSP is busy at award time. Another limitation is that the GRB manager has no obligation to inform potential contractors that an award has already been made. Sometimes, a manager may not receive bids for several reasons: (a) all potential GSPs are busy with other tasks, (b) a potential GSP is idle but ranks the proposed tender/task below the other tasks under consideration, (c) no GSPs, even if idle, are capable of offering service (e.g., resource is Windows NT-based, but user wants Linux). To handle such cases, a GRB can request quick response bids to which GSPs respond with messages such as *eligible, busy, ineligible* or *not interested*. This helps the GRB in making changes to its work plan. For example, the user can change deadline or budget to wait for new GSPs or attract existing GSPs to submit bids.

The tender (contract net protocol) model allows directed contracts to be issued without negotiation. The selected GSP responds with an *acceptanc*e or *refusal* of award. This capability can simplify the protocol and improve the efficiency of certain services.

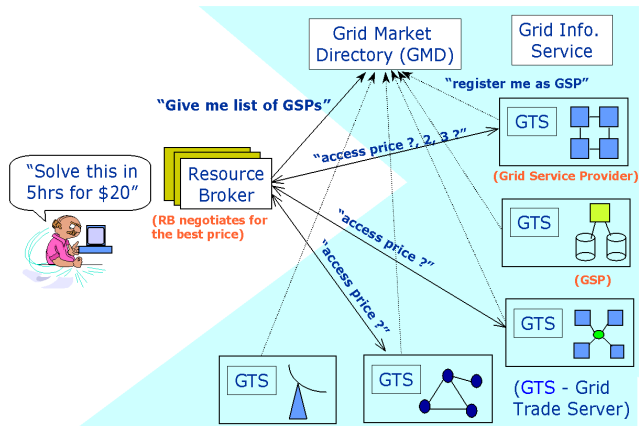
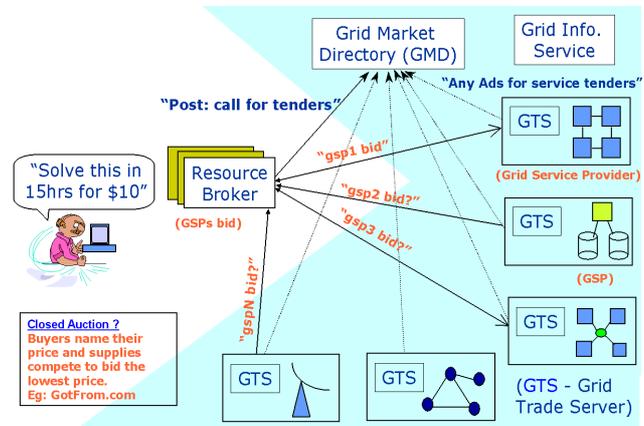

**Figure 3:** Brokers bargaining for lower access price in their bid for minimizing computational cost.

**Figure 4:** Tender/ContractNet model for resource trading.

## 3.5. Auction Model

The auction model supports one-to-many negotiation, between a service provider (seller) and many consumers (buyers), and reduces negotiation to a single value (i.e., price). The auctioneer sets the rules of auction, acceptable for the consumers and the providers. Auctions basically use market forces to negotiate a clearing price for the service.

In the real world, auctions are used extensively, particularly for selling goods/items within a set duration. The three key players involved in auctions are: resource owners, auctioneers (mediators), and buyers (see Figure 5). Many e-commerce portals such as Amazon.com and eBay.com are serving as mediators (auctioneers). Both buyers' and sellers' roles can also be automated. In a Grid environment, providers can use an auction protocol for deciding service value/price. The steps involved in the auction process are:
a. GSPs announce their services and invite bids.
b. Brokers offer their bids (and they can see what other consumers offer if they like - depending on open/closed).
c. Step (b) goes on until no one is willing to bid higher price or auctioneer may stop if minimum price line is not reached or owner's any other specific requirements are not meet.
d. GSP offers service to the one who wins
e. Consumer uses the resource

Auctions can be conducted as open or closed depending on whether they allow back-and-forth offers and counter offers. The consumer may update the bid and the provider may update the offered sale price. Depending on these parameters, auctions can be classified into four types [15]:

- English Auction (first-price open cry)
- First-price sealed-bid auction
- Vickrey (Second-price sealed-bid) auction [9]
- Dutch Auction

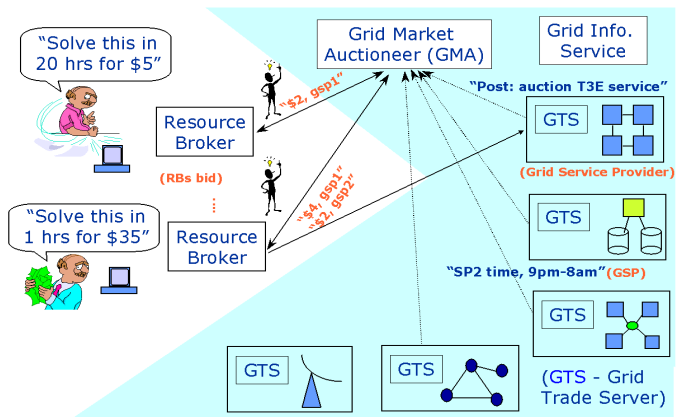 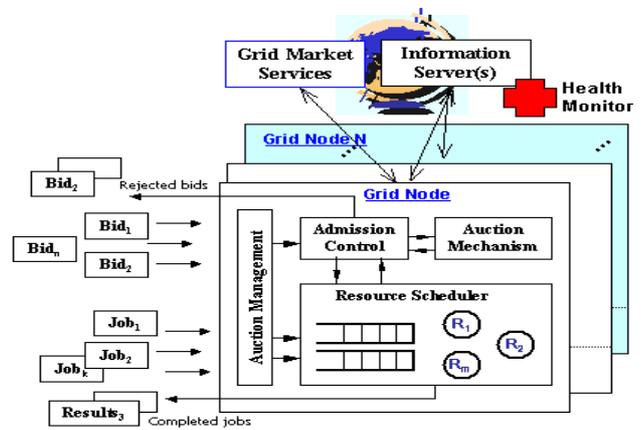

**Figure 5:** Auctions using external auctioneer.       **Figure 6:** Auction when GSPs use their own Auctioneer.

English Auction (first-price open cry) – all bidders are free to increase their bids exceeding others offers. When none of the bidders are willing to raise the price anymore, the auction ends, and the highest bidder wins the item at the price of his bid. In this model, the key issue is how GRBs decide how much to bid. A GRB has a private value (as defined by the user) and can have a strategy for a series of bids as a function of its private value and prior estimation of other bidder's valuations, and the past bids of others. The GRB decides the private value depending on the user-defined requirements (mainly deadline and budget that he is willing to invest for solving the problem). In the case of private value English auctions, a GRB's dominant strategy is to always bid a small amount "higher" than the current highest bid, and stop when its private value price is reached. In correlated value auctions, the policies are different and allow the auctioneer to increase the price a constant rate or at the rate he wishes. Those not interested in bidding anymore can openly declare so (open-exit) without re-entry possibility. This information helps other bidders and gives a change to adjust their valuation.

First-price sealed-bid auction – each bidder submits one bid without knowing the others' bids. The highest bidder wins the item at the price of his bid. In this case a broker bid strategy is a function of the private value and the prior beliefs of other bidders' valuations. The best strategy is bid less than its true valuation and it might still win the bid, but it all depends on what the others bid.

Vickrey (Second-price sealed-bid) auction— each bidder submits one bid without knowing the others' bids. The highest bidder wins the item at the price of the second highest bidder [9].

Dutch Auction – the auctioneer starts with a high bid/price and continuously lowers the price until one of the bidders takes the item at the current price. It is similar to first-price sealed-bid auction because in both cases bid matters only if it is the highest, and no relevant information is revealed during the auction process. From the broker's bidding strategic point of view, Dutch auction is similar to first-price sealed-bid auction. The interaction protocols for Dutch auction are as follows: the auction attempts to find market price for a good by starting a price much higher than the expected market value, then progressively reducing the price until one of the buyers accepts the price. The rate of reduction price is up to the auctioneer and they have a reverse price below which not to go. If the auction reduces the price to reverse price with no buyers, the auction terminates.

In terms of real time, Dutch auctions are efficient as the auctioneer can decrease the price at a strategic rate and first higher bidder wins. In an Internet wide auction, it is appealing in terms of automating the process wherein all parties can define their strategies for agents can participate in multiple auctions to optimize their objective function.

These auctions mainly differ in terms of whether they are performed as open or closed auctions and the offer price for the highest bidder. In open auctions, bidding agents can know bid value of other agents and will have an opportunity to offer next competitive bids. Whereas in closed auctions, bid offered by participants is not disclosed to others. Auctions can suffer from collusion (if bidders coordinate their bid prices so that the bids stay artificially low), lying auctioneers in case of Vickrey auction (auctioneer may overstate the second highest bid to the highest bidder unless that bidder can vary it), lying bidders, counter speculation, etc.

## *3.6.* *Bid-based Proportional Resource Sharing Model*

Market-based proportional resource sharing systems are quite popular in cooperative problem-solving environments like clusters (in single administrative domain). In this model, the percentage of resource share allocated to the user application is proportional to the bid value in comparison to other users' bids. The users are allocated credits or tokens, which they can use for having access to resources. The value of each credit depends on the resource demand and the value that other users place on the resource at the time of usage. For example, consider two users wishing to access a resource with similar requirements, but first user is willing to spend 2 tokens and the second user is willing to spend 4 tokens. In this case, the first user gets 1/3 of resource share whereas the second user gets 2/3 of resource share, which is proposal to the value that both users place on the resource for executing their applications.

This can be good way of managing a large shared resource in an organization or resource owned by multiple individuals (like multiple departments in a university) can have credit allocation mechanism depending on investment (or some policy). They can specify how much of credit they are willing to offer for running their applications on the resource. For example, a user might specify low credits for non-interactive batch jobs and high credits for interactive jobs with high response times. GSPs can employ this model for offering a QoS for higher price paying customers in a shared resource environment (as shown in Figure 7). Systems such as Rexec/Anemone and Xenoservers, D'Agents CPU market employ proportional resource sharing model in managing resource allocations [5].

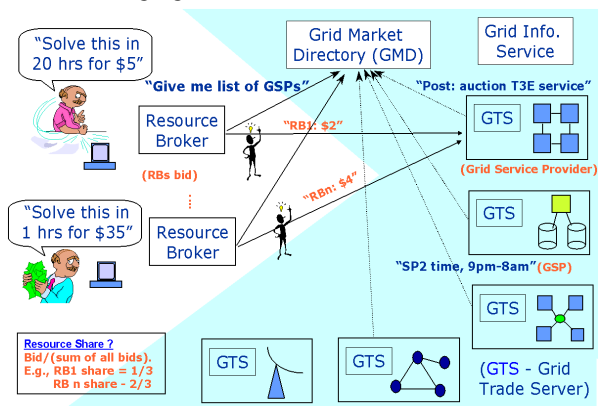

**Figure 7:** Market-based Proportional Resource Sharing.

## *3.7.* *Community/Coalition/Bartering/Share Holders Model*

A community of individuals shares each other's resources to create a cooperative computing environment. Those who are contributing their resources to a common pool can get access to that pool. A sophisticated model can also be employed here for deciding how much resources share contributors can get. It can involve credits that one can earn by sharing resource, which can them be used when needed. A system like Mojonation.net employs this model for storage sharing. This model works when those participating in the Grid have to be both service providers and consumers.

### *3.8. Monopoly/Oligopoly*

In the previously mentioned models we have assumed a competitive market where several GSPs and brokers/consumers determine the market price. However, there exist cases where a single GSP dominates the market and is the single provider of a particular service. In economic theory this model is known as a monopoly. Users cannot influence the prices of services and have to choose the service at the price given by the single GSP who monopolized the Grid marketplace. As regards the technical realization of this model, the single site puts the prices into the GMD or information services and brokers consult it without any possibility to negotiate prices.

Competitive markets are one extreme and monopolies are the other extreme. In most of the cases, the market situation is somewhere between these two extreme cases: a small number of GSPs dominate the market and set the prices. This market is known as an oligopoly.

### *3.9. Other Influences on Market Prices*

We now state more influences on price setting strategies in competitive, international markets. Supply and demand is the most common one but one also has to take into account national boarders and different pricing policies within different countries such as taxation, consumer price index, inflation, etc. These factors are not dealt in this paper, however implementations may need to consider them. There are micro and macro-economic factors that play an important role. Sure, one can also neglect them and build a price model on which all the Grid consumers have to agree. So this would correspond to an international market with special rules. Then, a model has to be formed for price changes. What is the factor for that change? Is there a monopoly that can decide what to do? Is the market transparent with optimally adapted prices etc. These are some of the main questions that need to be answered by a GSP when they decide their prices in an international market. A broker may consult the Grid Information Service to find out where the price for a particular service is minimal. For instance, one country might impose special taxes on a service whereas another country does not.

## 4. Economy in a Data Grid Environment

In computational Grid environments, large computational tasks that do not use very large amounts of data are solved. In Data Grid environments [10], large amounts of data are distributed and replicated to several sites all around the globe. Here, efficient access to the data is more important than scheduling computational tasks. When accessing large data stores, the cost for accessing data is important [12]. Can a single user afford to access a third of all data in a Petabyte data store? Certain restrictions and cost functions need to be imposed in order to obtain a good throughput for data access and to provide a fair response time for multiple users. An economic model can help achieve local or global efficiency. Paying higher prices can result in accessing a larger amount of data. How these prices can be mapped to the requirements of scientific users is still an open research issue. However, it is clear that some optimizations and restrictions for data access are required.

### *4.1. A Case for Economy in a Scientific Data Grid Environment*

In this subsection we discuss the possible use of economy in a scientific Data Grid environment, in particular in the DataGrid project [10]. We claim that scheduling based on economic models is different in a conventional "business environment" and in the scientific community. Whereas in the first, a Grid user pays explicitly a certain amount of money in order to get a service (e.g. pay Euro 100 for running application x in 10 minutes somewhere in the Grid), in the scientific community an explicit payment does not seem to be useful.

As regards the DataGrid project, several possible applications can be identified that require scheduling. Here, we concentrate only on the scheduling of user requests to (replicated) data. Let us define the problem. We assume several concurrent users (in the order of 100), replicated data stores with several Terabytes, up to Petabytes of data each and a finite throughput of data servers at a single site. A site can have several data servers and a hierarchical disk pool with tapes but each single site will have a restriction on the maximum amount of data it can serve at a time. Thus, an optimization problem occurs that tries to optimize the throughput for a large user community. For instance, a single user should not be able to request a Terabyte of data per day and consequently use all the data server resources. A fair scheduling concept is required that allows multiple, concurrent user requests. The problem can be identified as a high throughput problem. A scheduler at a single site has to take care of a fair scheduling concept. This compares to the conventional market model where a single user can increase his own throughput by offering to pay higher prices for a service. In the Data Grid environment a single user is interested in analyzing data and does not want to pay money explicitly for the physics analysis job. In other words, data analysis should not be restricted by resource prices since the main focus of physics analysis is to find some new information in large amounts of data rather than making money by selling data resources. However, this needs to be regulated to provide fair and improved quality of service to all users.

Data access in a Data Grid environment needs to be measured, regulated, and it has to be translated into costs, but shall not be expressed in currency units. One possible mapping of access requests to costs is to define the maximum data throughput of a site in the Data Grid. Based on this, a maximum number of tokens can be distributed to the users. For instance, a distributed data server at a site is able to serve 10 TB of data per day and thus 10 000 000 tokens are available and distributed to possible users. By default, a single user may only access as much data as he has tokens. This gives other users a chance to access data. However, the amount of data that they access for a given token need to be based on parameters such as demand, system load, QoS requested, etc. This helps in better managing resources and improving QoS offered by the system. The users can tradeoff between QoS and tokens.

The local site scheduler has to take care of admission control. Now, a single user might want to access more data than the available tokens. This requires re-distribution and negotiation of tokes. The negotiation step can then be done with the various economic models discussed earlier. Each user gets a certain amount of budget (expressed in tokens) and can be expressed in terms of pre-allocation priorities. For instance, users A and B are allocated X and Y number of tokens. Tokens can be renewed each day based on the current demand and the data server capabilities for serving data. The tokens can be converted into money when we compare it to an commercial environment (like 10 tokens for $1). When users want to use data services in the Data Grid, the system says, "During Peak time, I will charge 10 tokens per MB of data access and during off-pear time I will charge 6 tokens per MB." Accessing replicas from different sites may also result in having different prices for different replicas. If this is case, this itself is an economy. Consequently, tokens are assigned to users depending on the value/importance of users and their work.

We conclude that in a scientific Data Grid environment economic models can be applied, but a scheduling instance is still required that controls the overall throughput for a large user community.

## *4.2. Data Economy*

In [12] a cost model for distributed and replicated data over a wide area network is presented. Cost factors for the model are the network, data server and application specific costs. Furthermore, the problem of job execution is discussed under the viewpoint of sending the job to the required data (code mobility) or sending data to a local site and executing the job locally (data mobility). In the economic model, the main focus is on executing a job at any site as long as the cost for job execution is minimal.

Based on the previous case study we summarize this topic by intruding the term *Data Economy* that applies to several fields where data is distributed over Grids. The domains include

- Content (like music, books, magazines, news papers) sales using techniques such as micro-payments, aggregation, subscription, and subsidy.
- Community/Bartering model for content sharing
- Data and replica sites access in Data Grid environments

# 5. Scheduling Experiments on the World Wide Grid Testbed

In our work, for a given deadline and budget we have performed scheduling experiments at two different times (Australian peak and off-peak hours) on resources distributed in two major time zones [5] using "Cost-Optimization Scheduling algorithm" [4] on the World Wide Grid (WWG) [16] testbed (formerly called the Intercontinental Grid [5]). Currently, the WWG testbed has heterogeneous computational resources distributed across five continents: Asia, Australia, Europe, North America, and South America.

We performed an experiment of 165 CPU-intensive jobs, each running approximately 5 minutes duration. For a given deadline of 2 hours (120 minute) and budget of 396000 (Grid currency, G$, we can think of this as tokens/units that can be exchanged with real money), conducted experiments for two different optimization strategies:

1. Optimize for Time (produce results as early as possible, but before deadline and be within budget limit)

2. Optimize for Cost (produce results by deadline, but reduce cost and be within budget limit).

In these scheduling experiments, our Nimrod/G resource broker employed the *Commodity Market* model for establishing a service access price. It used Grid resource trading services for establishing connection with the Grid Trader running on GSP machines and obtained service prices. Our broker architecture is generic enough to use any protocol (including Tender/ContractNet or Auctions) for negotiating access to resources and choosing appropriate ones. The access price varies from one consumer to another consumer and from time to time as defined by resource owners. Depending on the deadline and the specified budget, the broker comes out with a plan for assigning jobs to resources. While doing so, it takes the capability of resources and the amount of resource share that the consumer can command (get) on that resource, and does load profiling to dynamically learn the ability of resources for executing jobs. It adapts itself to the changing resource conditions including failure of resources or jobs on the resource. The heuristics-based scheduling algorithms employed by Nimrod/G broker are presented in our early work [4].

The six computational resources in the World Wide Grid (WWG) [16] testbed have been selected in the economics driven scheduling experimentations. Table 1 shows resources details such as architecture, location, and access price along with type of Grid middleware systems used in making them Grid enabled. These are shared resources and hence they were not fully available to us. The access price indicated in the table is being established dynamically using GRACE resource trading protocols (commodity market model).

| Resource Type & Size (No. of nodes) | Organization & Location | Grid Services and Fabric | Price (G$ per CPU sec.) | Jobs Executed on Resources | |
|---|---|---|---|---|---|
| | | | | Time_Opt | Cost_Opt |
| Linux cluster (60 nodes) | Monash, Australia | Globus, GTS, Condor | 2 | 64 | 153 |
| Solaris (Ultra-2) | Tokyo Institute of Technology, Japan. | Globus, GTS, Fork | 3 | 9 | 1 |
| Linux PC (Prosecco) | CNUCE, Pisa, Italy | Globus, GTS, Fork | 3 | 7 | 1 |
| Linux PC (Barbera) | CNUCE, Pisa, Italy | Globus, GTS, Fork | 4 | 6 | 1 |
| Sun (8 nodes) | ANL, Chicago, USA | Globus, GTS, Fork | 7 | 42 | 4 |
| SGI (10 nodes) | ISI, Los Angeles, USA | Globus, GTS, Fork | 8 | 37 | 5 |
| | | Total Experiment Cost (G$) | | 237000 | 115200 |
| | | Time to Complete Experiment (in Min.) | | 70 | 119 |

**Table 1**: The WWG testbed resources used in scheduling experiments, Job execution and costing.

The number of jobs in execution on resources (Y-axis) at different times (X-axis) during the experimentation is shown in Figure 8 and Figure 9 for Time and Cost Optimization strategies respectively. In the first (Time Minimization) experiment, the broker selected resources in such a way that the whole application execution is completed at the earliest for a given budget. It completed execution of all jobs within *70 minutes* and spent *237000 G$*. In the second experiment (Cost Minimization), the broker selected cheap resources as much as possible to minimize the execution cost and yet meet the deadline (completed in *119 minutes*) and spent *115200 G$*. After the initial *calibration phase*, the jobs were distributed to the cheapest machines for the remainder of the experiment. The cost for completing time optimization experimentation is much larger than cost optimization due to the usage of expensive resources to complete the experiment early and at the same time ensure that it is within a specific budget. Interesting, the resources used in executing jobs are located on four different continents: Australia (Melbourne), Asia (Tokyo), Europe (Italy), North America (USA). The results show that our Grid brokering system can take advantage of economic models and user input parameters to meet their requirements. This provides consumers the ability to tradeoff between timeframe and cost that they would like to invest for solving the problem in hand. When the deadline is too tight and results are needed at the earliest possible time, then consumers should be prepared to spend more money.

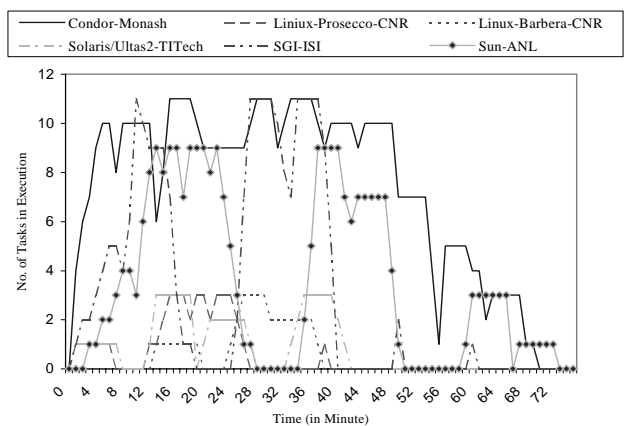

**Figure 8:** Resource Selection in Deadline and Budget based Scheduling for Time Optimization strategy.

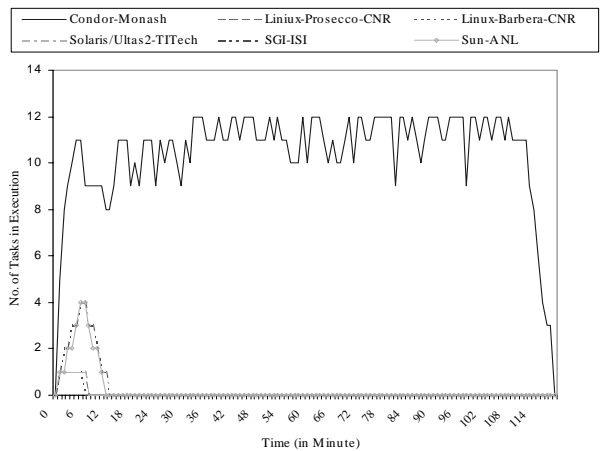

**Figure 9:** Resource Selection in Deadline and Budget based Scheduling for Cost Optimization strategy.

## 6. Conclusion and Future Work

Peer-to-peer and Grid computing technologies are enabling the creation of virtual enterprises (VE) for sharing distributed resources and changing the way we compute, communicate, and interact with systems and people. On the fly creation of Internet-scale virtual computing environments is becoming more of a reality than dream. The systems managing resources in this complex environment need to be smart, adaptable to changes in the environment and user requirements. At the same time, they need to provide a scalable, controllable, measurable, and understandable policy for management of resources. To meet all these requirements, we proposed economics paradigm for resource management and scheduling in P2P and Grid computing environment. We discussed several market-based economic models such as commodity market, tenders, and auctions that can be used for regulating the supply and demand in Grid-based virtual enterprises. We have demonstrated the power of these economic models in scheduling computations using the Nimrod/G resource broker on the World Wide Grid testbed spanning across five continents. This approach provides incentives for resource owners to share their resources on the Grid. More importantly, it provides mechanisms to trade-off QoS parameters, deadline and computational cost and incentive for relaxing their requirements—reduced computational cost if deadline is relaxed when timeframe for earliest results delivery is too critical.

First experiments that span several continents show promising results and provide a good basis for further work on Grid scheduling based on economics. We are carrying out simulations to investigate scalability, behavior, quantification, and applicability of economic models and scheduling algorithms in managing resources given very large Grid environment. We are currently implementing a new millennium version of Nimrod/G broker to provide highly programmable task-farming machinery with plug-gable components for scheduling algorithms, actuators, Grid agents, and computational steering modules. Efforts are underway to provide a clean interface to encourage problem solving environments and applications developers to use Nimrod/G brokering services in solving their problems in the Grid.

## References

[1] D. Abramson, J. Giddy, and L. Kotler, *High Performance Parametric Modeling with Nimrod/G: Killer Application for the Global Grid?*, IPDPS'2000, Mexico, IEEE CS Press, USA, 2000.
[2] R. Buyya, D. Abramson, and J. Giddy, *Nimrod/G: An Architecture for a Resource Management and Scheduling System in a Global Computational Grid*, HPC ASIA'2000, China, IEEE CS Press, USA, 2000.

[3] R. Buyya, D. Abramson, J. Giddy, *An Economy Driven Resource Management Architecture for Global Computational Power Grids*, The 2000 International Conference on Parallel and Distributed Processing Techniques and Applications (PDPTA 2000), Las Vegas, USA, June 26-29, 2000.

[4] R. Buyya, J. Giddy, D. Abramson, *An Evaluation of Economy-based Resource Trading and Scheduling on Computational Power Grids for Parameter Sweep Applications*, The Second Workshop on Active Middleware Services (AMS 2000), In conjunction with HPDC 2001, August 1, 2000, Pittsburgh, USA (Kluwer Academic Press).

[5] R. Buyya, D. Abramson, and J. Giddy, *An Economy Grid Architecture for Service-Oriented Grid Computing*, 10th IEEE International Heterogeneous Computing Workshop (HCW 2001), with IPDPS 2001, SF, California, USA, April 2001.

[6] K. Czajkowski, I. Foster, N. Karonis, C. Kesselman, S. Martin, W. Smith, and S. Tuecke, *A Resource Management Architecture for Metacomputing Systems,* IPPS/SPDP '98 Workshop on Job Scheduling Strategies for Parallel Processing, 1998.

[7] S. Chapin, J. Karpovich, A. Grimshaw, *The Legion Resource Management System*, Proceedings of the 5th Workshop on Job Scheduling Strategies for Parallel Processing, April 1999. (http://legion.virginia.edu/ )

[8] M. Stonebraker, R. Devine, M. Kornacker, W. Litwin, A. Pfeffer, A. Sah, C. Staelin, *An Economic Paradigm for Query Processing and Data Migration in Mariposa*, Proceedings of 3rd International Conference on Parallel and Distributed Information Systems, Austin, TX, USA, 28-30 Sept. 1994. Los Alamitos, CA, USA: IEEE Comput. Soc. Press, 1994.

[9] W. Vickrey, *Counter-speculation, auctions, and competitive sealed tenders*, Journal of Finance, Vol. 16, 1961.

[10] W. Hoschek, J. Jaen-Martinez, A. Samar, H. Stockinger, and K. Stockinger, *Data Management in an International Data Grid Project*, Proceedings of the First IEEE/ACM International Workshop on Grid Computing (GRID 2000), Springer Verlag Press, Bangalore, India, Dec. 17, 2000.

[11] L. McKnight and J. Boroumand, *Pricing Internet Services: Approaches and Challenges*, IEEE Computer, Feb. 2000.

[12] H. Stockinger, K. Stockinger, E. Schikuta, I. Willers. *Towards a Cost Model for Distributed and Replicated Data Stores*, 9th Euromicro Workshop on Parallel and Distributed Processing (PDP 2001), IEEE CS Press, Italy, February 7-9, 2001.

[13] D. Ferguson, C. Nikolaou, J. Sairamesh, Y. Yemini, "Economic Models for Allocating Resources in Computer Systems", In *Market-based Control: A Paradigm for Distributed Resource Allocation*, World Scientific Press, Singapore, 1996.

[14] R. Smith and R. Davis, *The Contract Net Protocol: High Level Communication and Control in a Distributed Problem Solver*, IEEE Transactions on Computers, Vol. C-29, No. 12, Dec. 1980.

[15] T. Sandholm, *Distributed Rational Decision Making*, Multi-Agent Systems (G. Weiss, editor), The MIT Press, 2000.

[16] R. Buyya, *World Wide Grid testbed*, http://www.csse.monash.edu.au/~rajkumar/ecogrid/wwg/, June 2001.

[17] American Heritage, *Dictionary of the English Language*, http://www.dictionary.com/cgi-bin/dict.pl?term=economics, June 2001.

[18] G. Heiser, F. Lam, and S. Russell, *Resource Management in the Mungi Single-Address-Space Operating System,* Proceedings of Australasian Computer Science Conference, Perth Australia, Feb. 1998, Springer-Verlag, Singapore, 1998.

[19] Y. Amir, B. Awerbuch., A. Barak A., S. Borgstrom, and A. Keren, *An Opportunity Cost Approach for Job Assignment in a Scalable Computing Cluster*, IEEE Tran. Parallel and Distributed Systems, Vol. 11, No. 7, July 2000.

[20] C. Waldspurger, T. Hogg, B. Huberman, J. Kephart, and W. Stornetta, *Spawn: A Distributed Computational Economy*, IEEE Transactions on Software Engineering, Vol. 18, No. 2, Feb. 1992.

[21] N. Nisan, S. London, O. Regev, N. Camiel, *Globally Distributed computation over the Internet - The POPCORN project*, International Conference on Distributed Computing Systems (ICDCS'98), 1998.

[22] B. Chun and D. Culler, *Market-based proportional resource sharing for clusters*, Technical report, University of California, Berkeley, September 1999.

[23] SETI@Home - http://setiathome.ssl.berkeley.edu/, June 2001.

[24] Mojo Nation - http://www.mojonation.net/, June 2001.

[25] N. Kapadia and J. Fortes, *PUNCH: An Architecture for Web-Enabled Wide-Area Network-Computing*, Cluster Computing: The Journal of Networks, Software Tools and Applications, September 1999.

[26] F. Berman and R. Wolski, *The AppLeS Project: A Status Report*, Proceedings of the 8th NEC Research Symposium, Berlin, Germany, May 1997.

[27] H. Casanova and J. Dongarra, *NetSolve: A Network Server for Solving Computational Science Problems*, Intl. Journal of Supercomputing Applications and High Performance Computing, Vol. 11, Number 3, 1997.

[28] M. Litzkow, M. Livny, and M. Mutka, *Condor - A Hunter of Idle Workstations*, Proceedings of the 8th International Conference of Distributed Computing Systems, June 1988.